\documentclass[preprint,aps]{revtex4}
\usepackage{graphicx}
\begin{document}

\title{Adsorption of Xe and Ar on Quasicrystalline Al-Ni-Co}
\author{Raluca A. Trasca, Nicola Ferralis, Renee D. Diehl and Milton W.Cole}
\address{Department of Physics and Materials Research Institute, Pennsylvania
State University, University Park, PA 16802}

\begin{abstract}
An interaction potential energy between an adsorbate (Xe and Ar) and the 10-fold Al-Ni-Co 
quasicrystal is computed by summing 
over all adsorbate-substrate interatomic interactions. The quasicrystal atoms' coordinates 
are obtained 
from LEED experiments and the Lennard-Jones parameters of Xe-Al, Xe-Ni and 
Xe-Co are found using semiempirical combining rules. The resulting potential
energy function of position is highly corrugated.

Monolayer adsorption of Xe and Ar on the quasicrystal surface is investigated in two cases: 
1) in the limit of 
low coverage (Henry's law regime), and 2) at somewhat larger coverage, when interactions
between adatoms are considered through the second virial coefficient, $C_{AAS}$.
A comparison with adsorption
on a flat surface indicates that the corrugation enhances the effect of the Xe-Xe (Ar-Ar) 
interactions. The theoretical results for the low coverage adsorption regime
are compared to experimental (LEED isobar) data.
\end{abstract}

\maketitle

\section{Introduction}

The growth and equilibrium structure of an adsorbed film are dependent on
the competing adsorbate-adsorbate and adsorbate-substrate interactions. The
laterally aperiodic adsorption potential of a quasicrystalline (QC) surface
provides an interesting case of competing interactions for rare gas
adsorbates, which favor close-packed monolayer structures in the absence of
substrate corrugation. Such incongruity
often produces new phenomena which are interesting and exotic on a
fundamental level. In this case, because quasicrystals can have radically
different physical properties than their periodic counterparts, it also
provides a significant tool for the design and growth of thin films having
specific properties.

The 10-fold surface of decagonal Al-Ni-Co is aperiodic in the surface plane
but periodic in the perpendicular direction \cite{Steurer,cervellino,takakura,
saitoh,steinhardt,haibach,mihalk,henley,cockayne,abe,zaharko}. Each
quasicrystalline layer comprises an aperiodic array of Al, Ni and Co with
points of 5-fold rotational symmetry.  Each plane is related to its
neighboring planes by a rotation of 36 degrees, producing an ABAB stacking
sequence. The structure of this surface has been studied using various
techniques, including low-energy electron diffraction (LEED) and scanning
tunneling microscopy (STM). A combination of these techniques was used
recently to demonstrate that the surface structure of Al-Ni-Co is similar to
the bulk structure determined by x-ray diffraction \cite{Steurer}, but with a
significant relaxation of the top layer and some intralayer buckling \cite{Nicola}.

There have been many theoretical studies of physical adsorption on flat
substrates and on periodic substrates \cite {Milton,landoldt}, but none on aperiodic substrates.
Recently a rigid-lattice total energy calculation for Al adsorption on a
quasiperiodic substrate produced some interesting results concerning the
growth and size of Al clusters on the surface \cite{lusher}. That calculation
was carried out by assuming a Lennard-Jones (LJ) potential between the adsorbate
Al atoms and the substrate atoms, assuming a bulk-like structure for the
surface.  Comparisons were made to the results of LEED experiments for the
structure, orientation and domain size of the Al film.  In particular, the
domain size distribution was shown to be a direct consequence of the
competing interactions in the substrate and the film.

Because the interactions are weaker, simpler and better known for physisorbed gases, we
believe we can gain a fuller understanding of the effects of aperiodicity
and symmetry by studying rare gas adsorption on Al-Ni-Co.  In this study, we
have chosen Xe and Ar as the adsorbates. We have calculated the
gas-surface adsorption potential and the adsorption properties in the
low-coverage limit using a virial expansion.  The results of these
calculations are compared to the results for adsorption on periodic
substrates, and to thermodynamic LEED measurements of Xe adsorption on
quasicrystalline Al-Ni-Co.

\section {Experimental Characterization of the Al-Ni-Co Substrate and Xe
Adsorption}

The decagonal Al-Ni-Co quasicrystal has a basic structure that consists of 
a stack of identical or nearly-identical 5-fold symmetric planes, each
related to adjacent ones by a $\pi/5$ rotation \cite{cervellino}. This produces a stacking
structure (ABAB) of A and B terminations, with a 10-fold screw axis. A
schematic representation of the atomic positions of one of such planes is
shown in Figure 1, where the different chemical identities and different local
geometry among the same chemical identity are specified.  The 10-fold surface
presents a contraction (10$\%$) of the first layer and an expansion (5$\%$) of the
second layer, with a small degree of intralayer rumpling; in plane
reconstruction is minimal, if present. The two-dimensional atomic density of a
layer is 0.123 $\mathring{A}$ \cite{nicola2}. 

The quasicrystalline d-$Al_{73}Ni_{10}Co_{17}$(Co-rich phase) sample was 
grown at Ames National laboratories using the decantation method \cite{moog}. The
surface was obtained by a cut perpendicular to the 10-fold axis, and then
polished as described elsewhere \cite{nicola2} to obtain a surface that was within 
0.5  $^\circ$
of the ten-fold orientation. The sample preparation in ultra-high vacuum
consisted of several cycles of Ar ion bombardment (500 eV ions) for about 45
minutes, followed by annealing for six to eight hours at temperatures up to
1060 K, as measured by a K-type thermocouple in contact with the sample, and
an optical pyrometer. The LEED pattern after preparation was observed to have
well-defined spots and relatively low intensity between the primary spots, as
shown in Figure 2. The impurity level was below detectablility, as measured by
Auger electron spectroscopy. The symmetry of the LEED pattern is 10-fold, due
to the presence of two equivalent surface terminations rotated by 36 $^\circ$ \cite{renee1}.
This method of preparation has been shown to produce a surface having a
structure that is essentially identical to the bulk structure with the
exception of a small degree of surface relaxation \cite{Nicola}. The LEED intensities
of the diffraction spots were measured using a rear-view LEED system, with the
electron beam at normal incidence to the surface. The LEED adsorption isobars
were obtained by holding the Xe pressure at a fixed value while changing the
temperature and acquiring LEED frames. For each frame the integrated spot
intensity of a substrate diffraction peak was extracted and plotted versus the
acquisition temperature, giving the adsorption isobar \cite{nicola2}. The temperature
was measured using a chromel-alumel thermocouple in contact with the sample,
in a range between 60 and 140K. The coverage is assumed to be linearly related
to the attenuation of the intensity of non-specular substrate diffraction peak
in the submonolayer range, where one monolayer is defined by the break in the
isobar. Thus, the substrate peak intensity before adsorption is defined to
correspond to zero coverage, the intensity at the break in the isobar is
defined to be one monolayer, and the intermediate coverages vary inversely and
linearly with the intensity \cite{moog}.

\section{Calculation of the QC interaction potential}
 
The calculation of the interaction energy between Xe (or Ar) and the Al-Ni-Co quasicrystalline 
structure is performed by summing over 
the LJ pair interactions, Xe-Al, Xe-Ni and Xe-Co. In the following, we will refer to Xe as
the adsorbent, but a similar technique is used for Ar for which we will also present results.
Lusher and coworkers employed
a similar LJ superposition while exploring the growth of Al on
Al-Co-Ni QC surface \cite{lusher}.  They assumed that the interaction of Al with each
species of the QC is the same. Using this approximation, they were able to predict 
nanocrystal
growth of Al on the QC substrate. Here, we assume that there are two distinct pair
interactions: Xe-Al and Xe-TM, where TM stands for either transition metal, Co or Ni.

In order to find the Xe-Al and Xe-TM LJ parameters, we consider adsorption of Xe on separate 
elemental surfaces 
of Al(110) and Ni(100). The semiempirical arithmetic combining 
rules between a Xe gas ($\sigma_{Xe}=4.1 \, \mathring{A}$) and Al
or Ni crystals ($\sigma_{Al}=2.5 \, \mathring{A},\, \sigma_{Ni}=2.2 \,\mathring{A}$) yield 
$\sigma_{Xe-Al}= 3.3 \,  \mathring{A}$ and $\sigma_{Xe-TM} = 3.1\, \mathring{A}$.
In order to estimate the gas-solid well-depth $\epsilon$ parameters, we use the experimental
heats of adsorption of Xe on Al(110) ($Q_{Xe-Al}=190$ meV) and on Ni(100) ($Q_{Xe-Ni}=226$ meV)
\cite{landoldt}, which are basically a
measure of the interaction well depth, and the summed pairwise interaction of
Xe with Al(110) and Ni(100) crystals, respectively \cite{steele}.
Finally, we find $\epsilon_{Xe-Al} \simeq 24$ meV and $\epsilon_{Xe-TM} \simeq 23$ meV.
Notice that the LJ parameters of Xe-Al and Xe-TM are not very different. However, the
small difference in the parameter values and the small 
rumpling of the QC surface play a role in the adsorption potential, as we will show in the
following.

Once the Xe-Al and Xe-TM interactions are obtained, we sum them to produce the net Xe-QC
interaction. Before discussing the results, we note that they 
should be interpreted with caution since summation of pair potentials omits effects
of electron delocalization,
which have been found to occur on metallic substrates \cite{renee}.

As described in the introduction, the coordinates of the 
Al-Ni-Co QC atoms were found using LEED data \cite{Nicola}. Due to the QC's aperiodicity, periodic
boundary conditions are not appropriate and all substrate atoms have to be considered in
calculations. We simplify by selecting a computational cell that is large enough 
($56\ \mathring{A}\ \times \ 56\ \mathring{A}$) to describe the potential adequately.
One way to depict the potential energy $V(\bf{r})$ is to construct a ``minimum energy surface'',
defined as follows: for any surface-parallel position $\bf R$ =(x,y) we evaluate the position
z$(\bf{R})$=$z_{min}(\bf{R})$ at which $V(\bf r)$ is a minimum. The resulting well-depth and force constant
are called $D(\bf R)$ and $k(\bf R)$, respectively.

A contour plot of the minimum energy 
surface is shown in Figure 3(a), where the various curves represent isopotential surfaces.
The minimum energy surface exhibits the five-fold symmetry of the QC top layer. The
big (red) circles correspond to the repulsive regions of the minimum energy surface and
are placed on top of the Al-2 species (see Fig. 1). In general, the repulsive part of the minimum energy
surface follows closely the 
distribution of atoms of the top layer, with one exception: the TM atoms which are
buried deepest in the first layer. The insensitivity of the interaction energy to those atoms
is due to the fact that $\sigma_{Xe-TM}$ and  $\epsilon_{Xe-TM}$ are smaller than 
$\sigma_{Xe-Al}$ and $\epsilon_{Xe-Al}$.

The most striking feature of the
minimum energy surface is its high ``corrugation'': local potential minima
range from -150 meV to -270 meV. A corrugation of this order was also found on open 
crystalline structures such as reconstructed Si and Ge \cite{conrad,packard}, but is not 
usual for metallic surfaces. However, this corrugation does not have the meaning of 
the hopping barrier between adjacent sites. To better
illustrate this point, we plot the lateral variation of $V_{min}(x,y)$ along different path
in Figure 3 (b). The full and dashed curves correspond to vertical paths 
in Fig. 3(a) at x=-1.6 $\mathring{A}$ and x=1.6 $\mathring{A}$. 
The energy difference between local minima and bridges is very irregular
and ranges in general from few meV to $\approx$ 100 meV. We attribute this large range of local minima 
to the QC's
aperiodicity as well as the ``holes'' ( 4 $\mathring{A}$ width) of its lateral
structure \cite{Nicola}. A more complete analysis of the hopping barriers between
adjacent sites, with applications in diffusion, involves an identification of the 
adsorption sites and the saddle points between those. This 
investigation is in progress.

In order to illustrate the most attractive adsorption sites, in Figure 4 we plot 
the regions where the adsorption potential
has values between -270 meV and -220 meV. McGrath et al argued that quasicrystals could be potentially used as
templates for quasicrystalline 2D structures \cite{mcgrath1, mcgrath2}. With this in mind, they have studied 
adsorption of $C_{60}$ on the Al-Pd-Mn quasicrystal surface, which is known to exhibit depressions
of about 7 $\mathring{A}$ width in a pentagrid structure. STM images at low coverage
showed that $C_{60}$ molecules occupy some of those pentagonal holes. In a similar way,
Al-Ni-Co is a good candidate as a template for Xe or Ar. It is not clear from Figure 4 
if adsorption of Xe at low coverage would lead to a quasicrystalline 2D long-range order
structure. Monte Carlo simulations are in progress to investigate this.
The interaction
energy between Ar and QC shows very similar properties, the main difference being that
the local potential minima range from -60 meV to -130 meV.

Another way to depict this highly corrugated potential is provided in Figure 5(a). 
There is shown the lateral variation of $V({\bf r})$ in the x-z plane for constant z
and y=0 (see Figure 3 a). 
As expected, far from the surface (z=5.8 $\mathring{A}$), the potential
is high and nearly independent of x. Close to the surface (z=2.8 $\mathring{A}$), 
the potential exhibits
a large and irregular corrugation. An unexpected feature is the following: at some particular
values of x (x $\approx$ -25, - 15, -10, -2.5, 5, 20 $\mathring{A}$) the potential is
``more'' attractive close to the surface (z = 2.8 $\mathring{A}$), but is ``less'' 
attractive at distances
farther away from the surface (z = 3.8, 4.4 $\mathring{A}$). Recently, it 
has been argued that the corrugation can change sign even for an
elemental crystal \cite{jean}. 
This may lead
to a possible steering effect on adsorption on the QC, especially important in the
process of film growing. 

To better understand the interaction potential landscape,
we define a ``volume density of states'' f(V) as follows: f(V)dV equals the volume above
the QC surface such that the potential energy lies in the interval [V,V+dV]. Then
\begin{equation}
f(V)=\int d {\bf r} \, \delta [V-V({\bf r})]
=\int d {\bf R}\,\left \lbrace \frac {1}{|\partial V/\partial z|^{(1)}}+\frac {1}{|\partial V/\partial z|^{(2)}} \right \rbrace \, \theta [V+D({\bf R})]
\end{equation} 
where $\theta(x)$ is the unit step function.
In the second expression, the denominators equal the magnitudes of the surface-normal forces
$|\partial V/\partial z|^{(1)}$ and $|\partial V/\partial z|^{(2)}$ evaluated at the two
points (1,2) where $V=V(\bf r)$. Equation 1 has been numerically integrated and the result
appears as the full curve in Figure 5(b). Note that the full curve exhibits a broad maximum around
-150 meV, due to the heterogeneity of the substrate. We may understand this by considering
the behavior of f(V) in two limits,
low V and high V. The region of the potential energy minimum (low V) is particularly important.
At a given $(\bf{R})$, this may be evaluated with a local harmonic approximation :
$V({\bf r}) \simeq -D({\bf R}) + k({\bf R})[z-z_{min}({\bf R})]^2/2$ and
$|\partial V/\partial z|^{(1)} \simeq |\partial V/\partial z|^{(2)} = [2 k({\bf R}) (V+D({\bf R}))]^{1/2}$. Therefore,
f(V) becomes:
\begin{equation}
f_{har}(V)= \int d{\bf R} \, \theta[V+D({\bf R})] \sqrt{\frac{2}{k({\bf R})(V+D({\bf R}))}}
\end{equation}
The local harmonic approximation for the heterogeneous surface is shown as a dashed curve
in Figure 5(b). The agreement with the true f(V) is excellent at low energy, reproducing
the broad maximum due to the heterogeneity.

The harmonic approximation to f(V) can be further approximated by assuming that the behavior in the region of the potential 
energy minimum is the same for all ${\bf R}$, $D({\bf R})=D$ and 
$k({\bf R})=k$. This corresponds to a monolayer film on a flat surface, for which we choose
as well depth the average value $D_{ave}({\bf R})$=180 meV. In this case:
\begin{equation}
f_{flat}(V)= A \, \theta(V+D) \, \sqrt{\frac{2}{k(V+D)}}.
\end{equation}

Here A is the computational surface area.
Results of Equation 3 are shown in Figure 5(b) as the dotted curve. Notice that instead of a shoulder,
$f_{flat}(V)$ exhibits a singularity at $V=-D_{ave}$; manifestly, the smooth surface approximation
is not suitable for a QC.

At high energy, another approximation can be employed, which
is related to the attractive part of the LJ potential. For a semi-infinte substrate, at
positions far from the surface
$V(z) = -{C_3}/{z^3}$, where $C_3=(\pi /6) \sum_{i=1}^\gamma n_i C_6^{(i)}$ for a substrate
composed of $\gamma$ species of atoms, and $C_6^{(i)}=4 \epsilon_i \sigma_i^6$ is the coefficient of the attractive part
in the usual LJ potential. Using this potential in Equation 1, we obtain:
\begin{equation}
f_{high}(V)= \frac{A}{3} (\frac{C_3}{|V|^4})^{1/3}
\end{equation}
Equation 4 is shown as dot-dashed curve in Figure 5(b). The approximation employed
at high energy (dot-dash) does not match exactly the true f(V) (full curve) since it assumes a 
semi-infinite substrate (which is not the case of the QC used in calculations) and since
the volume considered in the numerical calculation extends to $z=\infty$ while in practice
the numerical calculation stopped at $z_{max}=10 \ \mathring{A}$.

\section{Adsorption at low coverage}

Adsorption at low coverages can be described in terms of a virial expansion \cite{steele, Milton}.
\begin{equation}
N_{excess}=k_H \beta P + C_{AAS} \beta^2 P^2 + ...
\end{equation}
The leading term of this expansion is all that is needed when the
coverage is so low that interactions between adsorbate atoms can be neglected. 
In this regime, called the Henry's law regime, the coverage is proportional to the
pressure (P) and the coefficient of proportionality is called the Henry's law constant:
\begin{equation}
k_H=\int d^3 {\bf r} [e^{-\beta V({\bf r})}-1].
\end{equation}
Here $\beta=1/(k_BT)$ and $V({\bf r})$ is the adsorbate-substrate interaction potential.
At somewhat higher coverage, the interactions between adatoms become important
and are taken into account through the second virial coefficient $C_{AAS}$:
\begin{equation}
C_{AAS}=\int d^3 {\bf r_1} \int d^3 {\bf r_2}  e^{-\beta [V({\bf r_1})+V({\bf r_2})]}[e^{-\beta u(|{\bf r_1-r_2}|)}-1]
\end{equation}
where $u(|{\bf r_1-r_2}|)$ is the interaction between two adatoms. While the Henry's law
coefficient involves a 3D integral which can be relatively easily calculated, the 
second virial coefficient is a 6D integral which is computationally expensive due to the
large domain of integration. 

Figure 6(a) presents an Arrhenius plot of
the coverage (or excess number per unit area) as a function of 1/T for Xe and Ar
in two regimes: very low coverage (Henry's law regime) and higher coverage 
($C_{AAS}$ included). At high T,
the contribution of the second virial term is negligible, whereas at low T, mutual
adatom interactions become important and the coverage is enhanced by the $C_{AAS}$ term. However,
when the second virial term greatly exceeds the first term, the virial expansion becomes 
divergent and its results cannot be trusted. Figure 6(b) presents a comparison of
our calculated results with an experimental isobar. The agreement between calculation and
experiment is rather good at low coverage (below $0.01 A^{-2}$) considering
the simplicity of the model.

An interesting question arises in the heterogeneous environment of the QC potential: how
does this heterogeneity affect the mutual adatoms' interactions?  To investigate this, 
we compute the second virial coefficient in the 2D approximation \cite{steele}:
\begin{equation}
B_{2D}=-\frac{C_{AAS}}{2k_H^2}
\end{equation}
On a substrate modelled as a flat 2D continuum, a monolayer film is perfectly mobile,
so $B_{2D}$ is not influenced by substrate but only by the
adatoms' interaction $u(r)$:
\begin{equation}
B_{2D}^{flat}=-1/2 \int dr [e^{-\beta u(r)}-1]
\end{equation}

Figure 7 presents a comparison between $B_{2D}$/area for a flat substrate and for the QC.
Note that the QC's corrugation enhances the effect of the adatoms' attraction since $B_{2D}$
is larger in magnitude for a QC than for a flat substrate. We believe that this is another consequence of the aperiodicity
and the semi-close packed structure of the QC. The increased attraction (also evident in the
isosteric heat) occurs because the corrugated adsorption potential tends to bring particles
closer together than a smooth surface. This effect is much stronger than in the case of a crystal,
where the regular corrugated potential enhances the interactions' effect to a smaller extent \cite{steele}.

The isosteric heat of adsorption can be computed from the equation
of state:
\begin{equation}
Q_{st}=-(\frac{d \ln P}{d\beta})_N
\end{equation}
In the Henry's law regime, the isosteric heat reduces to:
\begin{equation}
Q_{st}=k_B T-<V>
\end{equation}
where $<V>=(\int dr V e^{-\beta V})/(\int dr e^{-\beta V})$ is the mean value of the adsorption
potential. In the case of a flat substrate, assuming a harmonic adsorption potential,
$<V>=-D+k_B T/2$. Thus, the isosteric heat of a perfectly mobile
monolayer on a flat surface increases monotonically with T as $Q_{st}=D+k_B T/2$. For a heterogeneous 
substrate, however, the dependence of $Q_{st}$ on T must be determined numerically. Figure 8
presents the isosteric heat results for Xe and Ar in the Henry's law regime (coverage
independent) and including the second virial coefficient correction (coverage dependent).
Note that in both cases $Q_{st} \simeq D-k_B T/2$, meaning that 
$<V> \simeq -D+3 k_B T/2$; this occurs because Xe atoms tend to be confined three-dimensionally
on the QC (rather than 1D in the smooth surface mode). This finding is consistent with 
the very corrugated potential in Figure 3(b). A comparison
between the Henry's law regime and second virial coefficient correction 
shows that the effect of interactions is to increase the isosteric heat.
At T= 70 K, the calculated isosteric heat
of Xe is 290 meV in the Henry's law regime and 305 meV when interactions are included.
Both predictions are close to the value of
300 meV at 0.25 monolayer adsorption extracted from experimental isobars\cite{Nicola}.

\section{Conclusions}

In conclusion, we have computed the adsorption potential of Xe and Ar adsorbates
on Al-Ni-Co quasicrystalline substrate by adding gas-surface LJ interactions. 
Due to the substrate's aperiodicity and
heterogeneity, the adsorption potential was found to be highly corrugated, with local minima
ranging from -150 to -270 meV for Xe and -60 to -130 meV for Ar. The minimum energy surface
of the adsorption potential exhibits the 5-fold symmetry of the QC's top layer, with the 
most repulsive regions located on top of the QC's atoms and
the most attractive regions in the ``holes'' of the semi-close packed structure of
the QC. 

To understand the distribution of adsorption potential values, we have defined
a volume density of states f(V) and computed this quantity for QC, for a flat
surface and for a harmonic approximation of the interaction potential. While
$f_{flat}(V)$ exhibits a singularity when $V = -D$ (the well depth),
$f_{QC}(V)$ has a broad maximum due to the QC's ``heterogeneity''. This shoulder is well
reproduced by a local harmonic approximation of the adsorption potential in each point above
the QC surface.

To model adsorption at low coverages, a virial expansion of the equation of state was
employed, including first and second virial coefficient terms. Isobars obtained from the virial
expansion agree well with experimental isobars of Xe on Al-Ni-Co QC. The isosteric
heat of Xe was found to be 305 meV, close to experimental value of 300 meV at 0.25 monolayer. 
The dependence
of the isosteric heat on T indicates that adatoms are 3D-confined in the attractive
regions of the adsorption potential. This finding is consistent with the results of 
the second virial coefficient which suggest that the corrugation of the potential tends to
bring adatoms together and enhance their mutual interactions.
Future work will focus on a Monte Carlo study at arbitrarily
high coverage and on a search for the adsorbate ground state structure.

\section{Acknowledgments}

We thank L.W. Bruch and G.D. Mahan for edifying comments and the National Science Foundation 
(Grants 02-08520 and 03-03916) for support
of this research.

\newpage
\section{Figure captions}

1. One layer of a $45 \times 45\ \mathring{A}^2$ slab,
representing one 5-fold plane of the d-AlNiCo quasicrystal surface, along the
10-fold axis. The atom colors correspond to sublayer groups differentiated
according to the chemical identity and the local geometry (nearest neighbor
distance): Red - Transition Metals (Ni or Co) 1, Black - Transition Metals 2,
Green  Al-1, Blue  Al-2, Cyan  Al-3, Yellow  Al-4 \cite{Nicola}.

2. LEED pattern from the 10-fold surface of the
clean d-AlNiCo, acquired at 60 K. The incident energy was 55 eV.

3. Interaction potential energy between Xe (or Ar) and Al-Ni-Co QC:
	(a) Contour plot of the minimum energy surface. Different colors correspond to
	different isopotential curves: red = -175 meV, green = -200 meV, turqoise = -225 meV, 
	blue = -250 meV; (b) Lateral variation of the minimum potential along two paths:
	x= 1.6 $\mathring{A}$ (dashed curve) and x=-1.6 $\mathring{A}$ (full curve).

4. The preferred adsorption regions at low coverage (-270 meV $\leq \ V_{min}(r)\ \leq$ -220 meV).

5. Interaction potential energy between Xe and Al-Ni-Co QC:
	(a) Potential energy as a function of x for constant z above the QC, (b) The ``volume
	density of states'' of the QC (full curve), harmonic approximation limit at low energy (dashed), 
	the high energy limit approximation (dot-dashed) and for a flat surface (dotted).

6. Coverage as a function of temperature (a) Arrhenius plot of
	Xe and Ar isobars at P=1.6*$10^{-7}$ mbar
	in the Henry's law regime (Ar-full, Xe-dot) and 
including the second virial term (Ar-dot-dash, Xe-dash), (b) Comparison of calculations
with an experimental isobar of Xe at low coverage.

7. The second virial coefficient for a flat surface and the QC surface.

8. Isosteric heat of Xe and Ar on the QC in the Henry's law regime:
	Xe (dashed), Ar (dotted), and including the second virial coefficient correction:
	Xe (full), Ar (dot-dashed). 

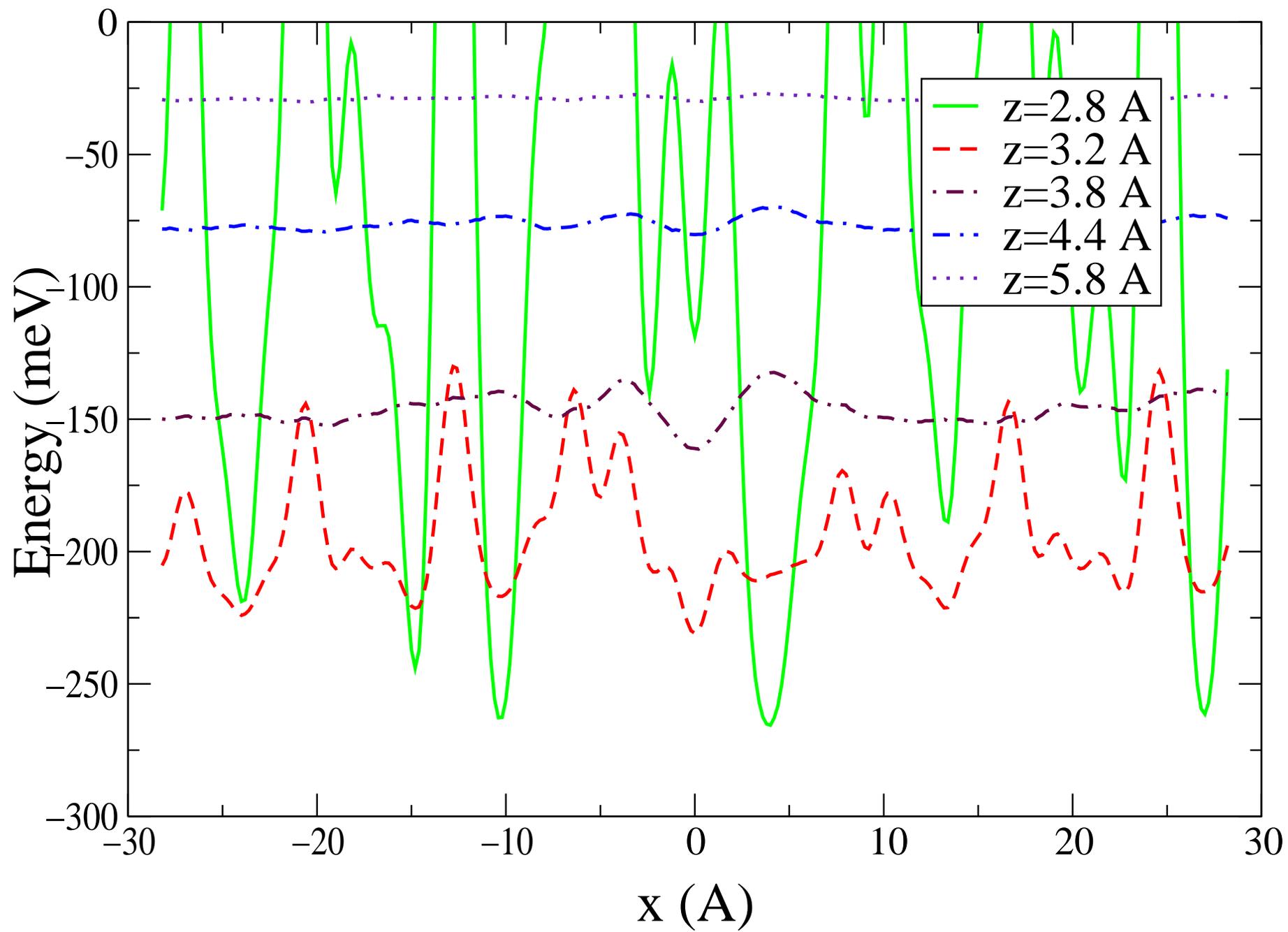
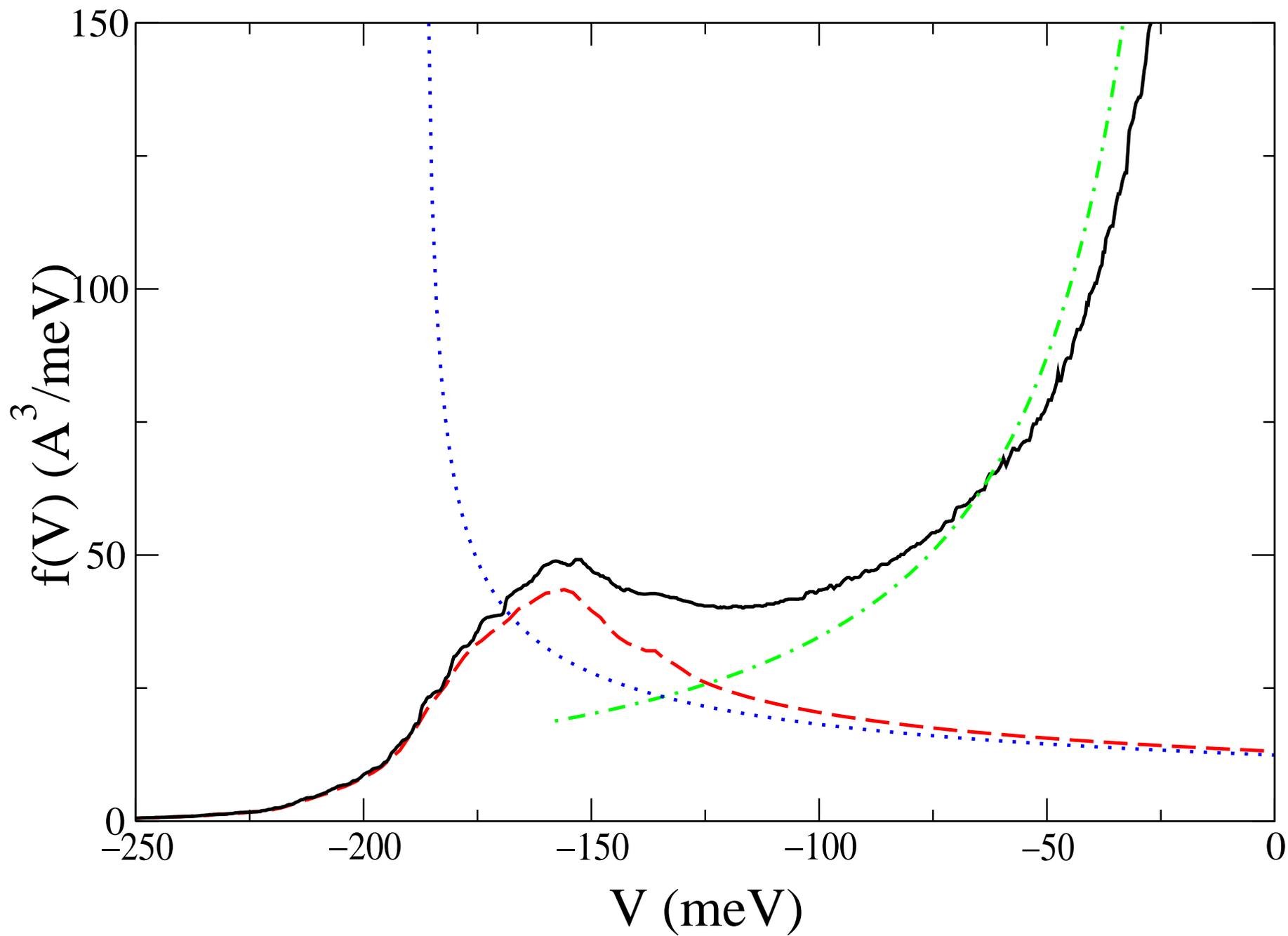
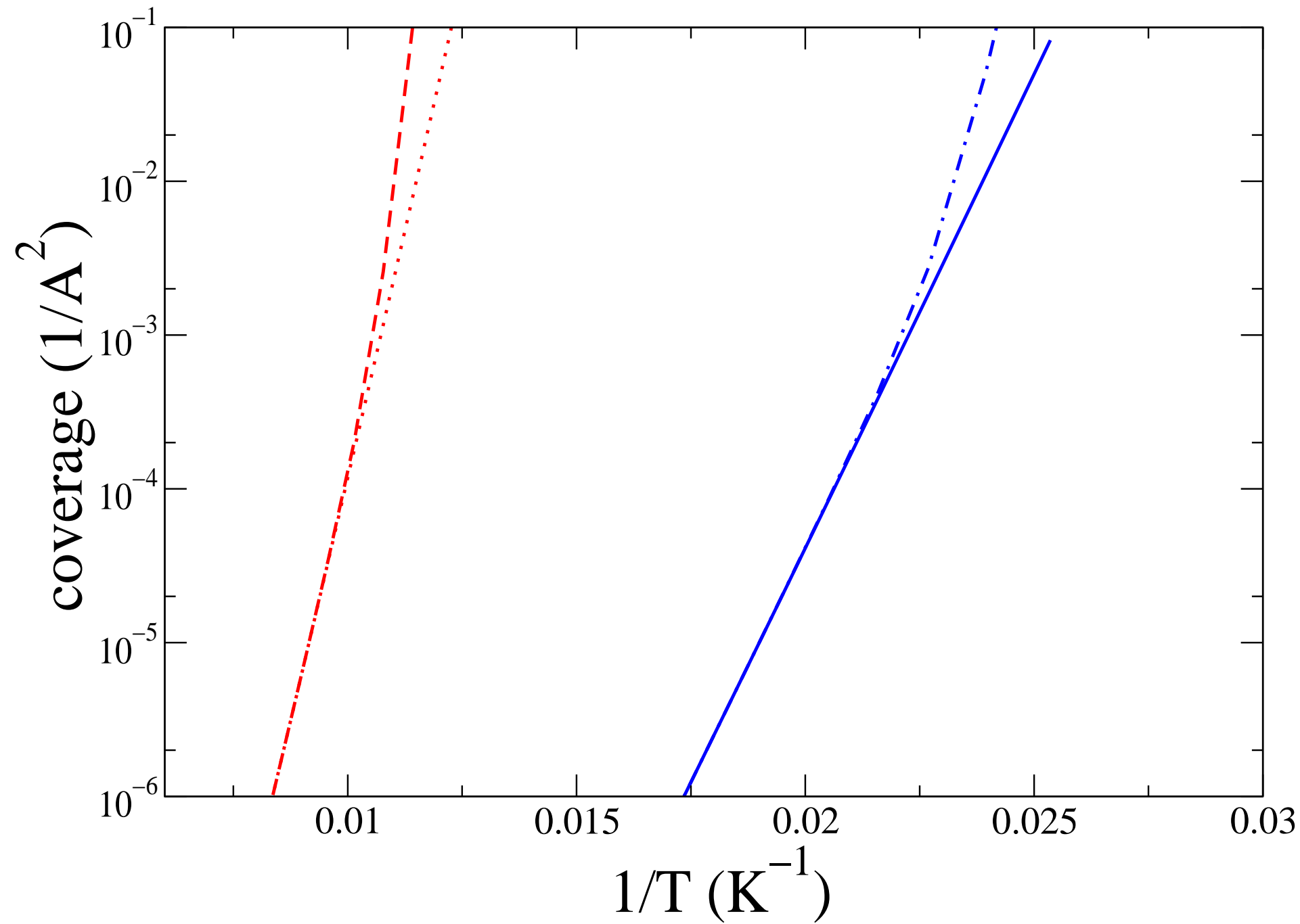
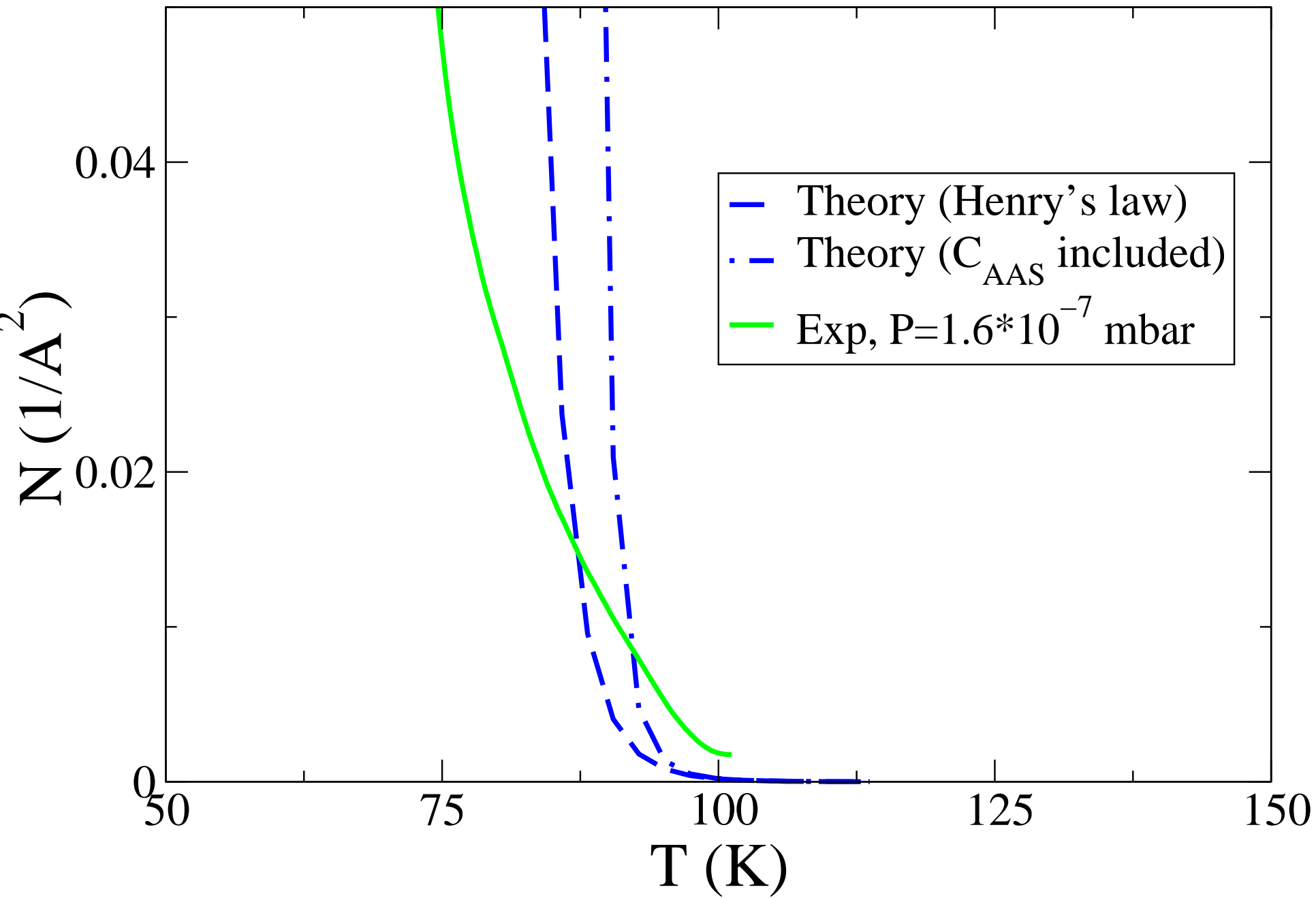
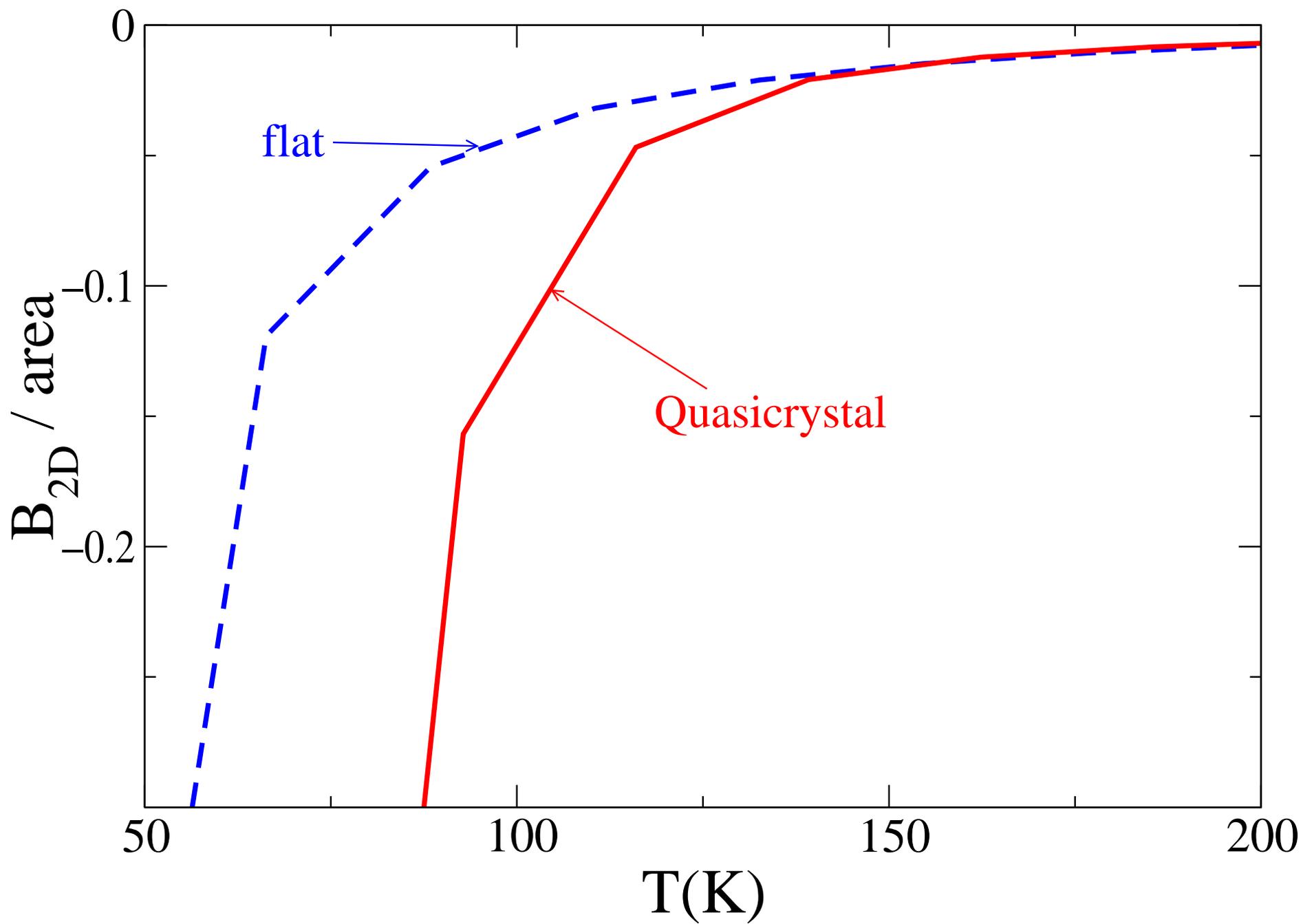
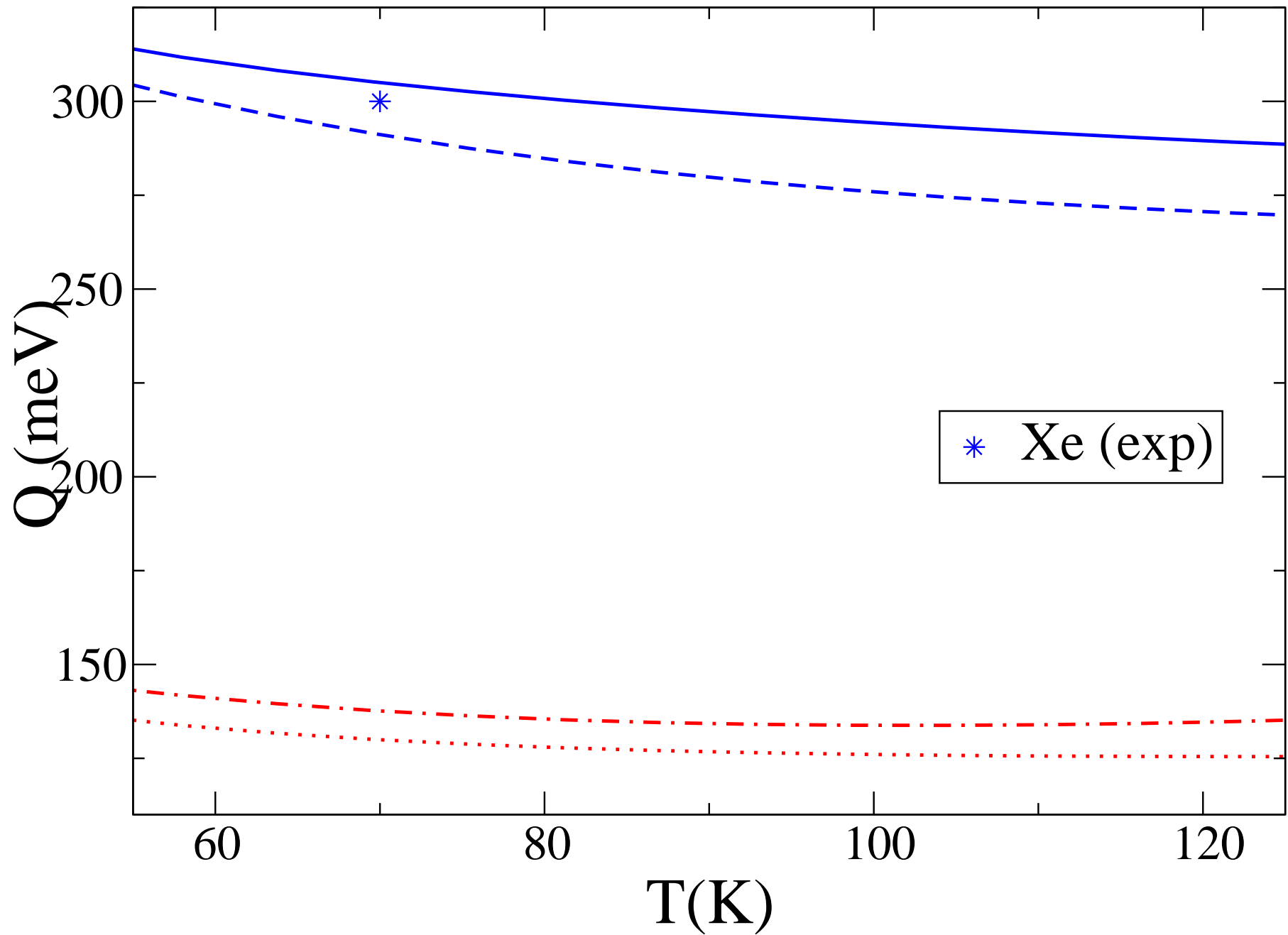
\end{document}